\documentclass[intlimits,sumlimits,12pt]{iopart}

\expandafter\let\csname equation*\endcsname\relax
\expandafter\let\csname endequation*\endcsname\relax

\newcommand{\abs}[1]{\ensuremath{\left\vert#1\right\vert}}
\newcommand{\vc}[1]{{\boldsymbol #1}}

\newcommand{\unit}{1\!\!1}

\renewcommand{\Re}{{\text{Re }}}
\renewcommand{\Im}{{\text{Im }}}

\usepackage{iopams}
\usepackage{amsmath}
\usepackage{dsfont}
\usepackage{color}
\usepackage{comment}
\usepackage{graphicx}
\usepackage[numbers]{natbib}

\usepackage{subfigure}
\usepackage{hyperref}

\begin{document}

\title[]{Winding Number Statistics of a Parametric Chiral Unitary Random Matrix Ensemble}

\vspace{1cm}

\hspace{-0.85cm}\textit{\large Dedicated to the Memory of Fritz Haake}

\vspace{0.2cm}

\author{Petr Braun$^{1}$\footnote{deceased}, Nico Hahn$^{1}$, Daniel Waltner$^{1}$, Omri Gat$^{2}$ and Thomas Guhr$^{1}$}

\address{$^{1}$ Fakult\"at f\"ur Physik, Universit\"at Duisburg--Essen, Duisburg, Germany\\
         $^{2}$ Racah Institute of Physics, The Hebrew University, Jerusalem, Israel}

\ead{nico.hahn@uni-due.de}


\begin{abstract}
The winding number is a concept in complex analysis which has, in the
presence of chiral symmetry, a physics interpretation as the
topological index belonging to gapped phases of fermions. We study
statistical properties of this topological quantity. To this end, we
set up a random matrix model for a chiral unitary system with a
parametric dependence.  We analytically calculate the discrete
probability distribution of the winding numbers, as well as the
parametric correlations functions of the winding number density.
Moreover, we address aspects of universality for the two-point
function of the winding number density by identifying a proper unfolding procedure. We conjecture
the unfolded two-point function to be universal.
\end{abstract}

%



\section{Introduction}
\label{sec0}

Chiral symmetry classes comprise three of the ten symmetry classes of
disordered fermions \cite{AltlandZirnbauer1997, Heinzner2005,
  Kitaev2009, Ryu2016, Oppermann1990}, also known as the tenfold way. They were
initially discovered in works on Quantum Chromodynamics (QCD) and
statistical properties of lattice gauge calculations
\cite{Verbaarschot1994}. The chiral symmetry of the Dirac operator is
broken spontaneously as well as explicitly by the quark masses. The
spectral properties of the Dirac operator are connected to the chiral
condensate, the order parameter of the phase transition that occurs at
a high temperature and that restores chiral symmetry. To the present
knowledge, simultaneously the confinement--deconfinement transition
takes place which frees the quarks by opening the hadronic particles.
The study of the chiral symmetry naturally establishes a link to
random matrix theory \cite{GuhrRMTReview, MehtaBook} and the theory of
disordered systems. The so emerging chiral random matrix theory turned
out to be a fruitful approach in low temperature quantum
chromodynamics \cite{VerbaarschotWettig2000,ShuryakVerbaarschot1993,
  Wettig1996A, Wettig1996B, JacksonVerbaarschot1996,
  VerbaarschotZahed1993,GWW2000}.  It is worth mentioning that
topological aspects play an important role in QCD, one object of
particular interest is the topological charge, a comparison of its
various definitions was recently given in Ref.~\cite{Alexandrou2020}.
However, these topological quantities do not seem to be directly
related to the ones we study in the present contribution.

In the context of condensed matter physics, however, chiral symmetry
may appear either as sublattice symmetry or as a combination of time
reversal and particle-hole symmetry \cite{Zirnbauer2021}. In an early
work \cite{Gade1993} localization in systems with such a sublattice
symmetry was observed. Here, the energy-level statistics at half
filling is different from the bulk statistics.  While the latter is
described by the random matrix theory of the classical Wigner--Dyson
ensembles, the former can be captured with chiral unitary, orthogonal,
or symplectic random matrices, depending on the additional presence of
time-reversal and spin-rotation symmetries or the lack thereof
\cite{Beenaker2015}.

Translationally invariant one-dimensional chiral systems that are
gapped at the center of the spectrum are also characterized by the
\emph{winding number}, an integer topological index associated with
the bundle of negative-energy bands. Systems with a nonzero winding
number $W$ are topologically nontrivial, and therefore have $\abs{W}$ modes
localized at each boundary \cite{ChenChiou2020, Shapiro2020}. An
intriguing example is the time-reversal invariant Majorana chain,
which belongs to the chiral orthogonal class BDI, and whose edge modes
have resilient quantum information properties \cite{Alicea2012}.

When (discrete) translation invariance is broken by disorder obeying
chiral symmetry, lattice momentum is no longer a good quantum
number. Nevertheless, it is possible to express $W$ in position
representation, and then calculate the winding number of a periodic
system; it is then found in weakly disordered one-dimensional systems
that the winding number is self--averaging and robust in the
thermodynamic limit \cite{Prodan2014, Altland2015}.

On the other hand, if the translation invariance is perturbed by
disorder that is itself periodic, translational invariance persists,
but with a possibly larger unit cell. In this case the winding number
is sensible also for strong disorder, as a random variable. The
probability distribution of $W$ in periodic systems with a disordered
unit cell depends on that of the disorder, but like the statistics of
energy levels, it may turn out that the winding number statistics
becomes universal when the unit cell becomes large, and moreover that
the universal distribution can be reproduced by random matrix models.

This question has not yet been addressed in the chiral classes, but
there are precursors in the unitary class A, where energy bands in two
dimensions are topologically classified by the (first) Chern
number. Random matrix models defined on compact two-dimensional
parameter spaces were studied in \cite{WalkerWilkinson1995,
  GatWilkinson2021}, showing that the Chern number distribution is
Gaussian with a universal covariance.

In \cite{GatWilkinson2021} the Chern number covariance was calculated
as an integral of the correlation function of the adiabatic curvature,
which is universal as well. The two-point correlation function of the
adiabatic curvature follows a scaling form, with a scale parameter
equal to the density of states multiplied by the correlation length of
the elements of the random matrices in parameter space, and a
universal scaling function. Universal scaling behaviour of this kind
has been known for a long time in parametric correlations of spectral
properties of random matrices, like the density and current of energy
states \cite{BeenakkerRejaei1994}. Furthermore, the universal
properties of parametric spectral correlations of random matrices
agree with those of disordered systems \cite{SimonsAltshuler1993A,
 SimonsAltshuler1993B}, motivating the universality hypothesis for
the correlations of the adiabatic curvature and its chiral class  analog, the winding number
density, and a fortiori the probability distributions of Chern numbers
and winding numbers.

We have the following goals: we want to calculate the discrete probability
distribution of the winding numbers as well as the first two moments.
We also wish to compute the parametric correlation functions
for the winding number densities. Furthermore, we discuss aspects
of universality for the two-point function by identifying an
unfolding procedure.

The paper is organized as follows: in Sec.~\ref{sec1} we introduce
chiral symmetry and the winding number. Furthermore we set up the
random matrix model and define the goals of this paper. In
Sec.~\ref{sec2} we present the main idea of our calculations and our
results. More involved derivations are relegated to Sec.~\ref{sec3}. We conclude in Sec.~\ref{sec4}.

\section{Posing the Problem}
\label{sec1}

We consider the chiral unitary symmetry class, labeled AIII in the
tenfold way \cite{Ryu2016, Kitaev2009, Schnyder2008,
  AltlandZirnbauer1997, Oppermann1990}.  We refer to the matrices in this class as
Hamiltonians $H$, even though they may represent Dirac operators in
the context of quantum chromodynamics.  Using the anticommutator
$\{,\}$, chiral symmetry can formally be expressed as
\begin{equation} \label{2ChiralSymmetry}
\{\mathcal{C},H\} = 0
\end{equation}
with the chiral operator $\mathcal{C}$. In our case, the matrices $H$
are complex Hermitean with dimension $2N\times 2N$, and the matrix
representation of the chiral operator reads in diagonal form
\begin{equation}
\mathcal{C} = \begin{bmatrix}
\unit_N & 0
\\
0 & -\unit_N
\end{bmatrix}.
\end{equation}
In the same basis a Hamiltonian obeying \eqref{2ChiralSymmetry} takes the block off--diagonal form
\begin{equation} \label{2ChiralHamiltonian}
H = \begin{bmatrix}
0 & K
\\
K^\dag & 0
\end{bmatrix},
\end{equation}
where the  $N\times N$ complex matrix $K$ has no symmetries. The chiral Gaussian Unitary
Ensemble (chGUE) consists of all these matrices with entries drawn from a Gaussian
probability distribution invariant under unitary rotations. Put differently, the matrices
$K$ form a complex Ginibre ensemble \cite{Ginibre1965}.

Topological properties can be explored by giving these random matrices
a parametric dependence $K=K(p)$ and thus $H=H(p)$, where the real
variable $p$ lies on a circle, i.e.~$p$ parametrizes the
one--dimensional manifold $\mathcal{S}^1$. The topological invariant associated with this class of
Hamiltonians is the winding number \cite{Maffei2018, AsbothBook}
\begin{equation} \label{2WindingNumberDef}
W = \frac{1}{2\pi i} \int_0^{2\pi} dp\ w(p),
\end{equation}
where
\begin{equation} \label{2WindingNumberDensityDef}
w(p) = \frac{d}{dp} \ln \det K(p) = \frac{1}{\det K(p)} \frac{d}{dp} \det K(p)
\end{equation}
is the winding number density. It is a standard result in complex analysis that the winding number is an
integer, $W\in \mathbb{Z}$, whenever $\det K$ is a nonzero analytic function of $p$.  To set up a concrete random matrix model for the chiral
Hamiltonians, we choose the parametric dependence in the explicit form
\begin{equation} \label{2RMF}
K(p) = K_1 \cos p + K_2 \sin p,
\end{equation}
where now the matrices $K_1$ and $K_2$ are $N\times N$ dimensional
complex matrices with independently Gaussian distributed elements,
just like $K$ in \eqref{2ChiralHamiltonian}. Hence, the sets of matrices $K_1$ and
$K_2$ form independent Ginibre ensembles. We denote an average over this
combined ensemble with angular brackets. The associated Hamiltonians
\begin{equation} \label{2RMFH}
H(p) = H_1 \cos p + H_2 \sin p \qquad \text{with} \qquad H_m = \begin{bmatrix}
0 & K_m
\\
K_m^\dag & 0
\end{bmatrix}, \  m=1,2 \ ,
\end{equation}
may thus be viewed as a defining a parametric combination of two
chGUE's. We also refer to $H=H(p)$ as a random matrix field.

We calculate the $k$-point correlation function of
winding number densities as a random matrix ensemble average, 
\begin{equation} \label{2kPointCorrelationDef}
C_k (p_1,\ldots,p_k) = \left\langle w(p_1) \cdots w(p_k) \right\rangle.
\end{equation}
The precise meaning of the angular brackets indicating the ensemble
average will be given in the sequel.  The arguments $p_i,
\ i=1,\ldots,k$ with $p_i \in [0,2\pi)$ are the different points on
  the parameter manifold.  Furthermore, we compute the
  distribution of winding numbers $P(W)$. An exact expression for its
  moments
\begin{equation} \label{2WindingNumberMomentsDef}
\left\langle W^k \right\rangle = \sum_{W \in \mathbb{Z}} W^k P(W) =
\frac{1}{(2\pi i)^k} \int_0^{2\pi} dp_1 \cdots \int_0^{2\pi}
dp_k\ C_k(p_1,\ldots,p_k)
\end{equation}
is given in terms of the $k$-point correlation function.

\section{Concepts and Results}
\label{sec2}

In Sec.~\ref{sec21} we sketch the strategy for our calculation of the
$k$-point correlation function, the details of the derivations are
collected in Sec.~\ref{sec3}. Quite remarkably, we arrive at
closed--form results for arbitrary $k$ and particularly simple
expressions for $k=1$ and $k=2$. In Sec.~\ref{sec21a}, we discuss
aspects of universality and unfold the parametric dependence of the two-point function. We conjecture the resulting limit to be universal. The
winding number distribution and its moments are addressed in
Sec.~\ref{sec22}.

\subsection{Expressions and Results for the $k$-Point Correlation Function}
\label{sec21}

For the specific form of our random matrix field \eqref{2RMF}, noting that $\det K_1\ne0$ with probability 1, we can
evaluate the logarithm appearing in Eq.~\eqref{2WindingNumberDensityDef}
as
\begin{align}
\ln \det K(p) &= \ln \det K_1 + N \ln \sin p + \ln \det \left( \cot p + K_1^{-1}K_2 \right)
\\
&= \ln \det K_1 + N \ln \sin p + \sum_{n=1}^N \ln \left( \cot p + z_n \right),	\nonumber
\end{align}
where  $z_n , \  n=1,\ldots,N$  are  the complex eigenvalues of  the  matrix
$K_1^{-1}K_2$, we  also use $z=(z_1,\dots,z_N)$. 
Taking  the derivative
yields the (unaveraged) winding number density
\begin{equation} \label{3WindingNumberDensity}
w(p) = N\cot p - \frac{1}{\sin^2 p}\sum_{n=1}^N \frac{1}{\cot p + z_n}.
\end{equation}
Here and below, intermediate singularities at $p=0,\pi$ cancel to yield analytic correlations functions for all values of $p$. The matrices $K_1^{-1}K_2$ form a so-called spherical ensemble \citep{Krishnapur2009}. The corresponding joint eigenvalue density is known,
\begin{align} \label{3SphericalEnsembleDistribution}
G(z) = G(z_1,\ldots,z_N) &= \frac{1}{c_N \pi^N} \abs{\Delta_N(z)}^2 \prod_{n=1}^N \frac{1}{(1+\abs{z_n}^2)^{N+1}}
\\
c_N &= N! \prod_{n=1}^N B(n,N-n+1),	\nonumber
\end{align}
where $B(n,m)$ is the Euler Beta function \cite{NIST} and
\begin{equation} \label{3VandermondeDeterminant}
\Delta_N(z) = \prod_{1\leq n<m\leq N} (z_n - z_m) = \det\left[ z_n^{m-1}\right]_{n,m=1,\ldots,N}.
\end{equation} 
is the Vandermonde determinant. With the volume element over the combined
$N$ complex planes
\begin{equation} \label{4VolumeElement}
d[z] = \prod_{i=1}^N d[z_n] \quad \text{where} \quad   d[z_n] = d\Re z_n\ d\Im z_n,
\end{equation}
we eventually arrive at a precise definition for the ensemble average of a function $F(z)=F(z_1,\ldots,z_N)$ as
\begin{align} \label{Eaverage}
  \langle F(z) \rangle &= \int d[z] G(z) F(z) \\
  &= \int  d[z_1] \cdots \int d[z_N] \, G(z_1,\ldots,z_N) \, F(z_1,\ldots,z_N).
\nonumber
\end{align}
In particular, for Eq.~\eqref{2kPointCorrelationDef}, we find that
\begin{equation} \label{Caverage}
C_k (p_1,\ldots,p_k) = \int d[z] \, G(z) \, w(p_1) \cdots w(p_k)
\end{equation}
is the integral we have to compute. For convenience, we suppress the
$z$ dependence in the argument of the function $w(p)$.

To proceed with the calculation of the integral \eqref{Caverage}, we
observe that the winding number density $w(p)$ according to
Eq.~\eqref{3WindingNumberDensity} features a term independent of the
eigenvalues $z_n$. We subtract this term by defining
\begin{align} \label{3yDef}
  y(p) &= w(p) - N q = -\frac{1}{\sin^2 p} \sum_{n=1}^N \frac{1}{q + z_n}\\
     q &= \cot p .	\nonumber
\end{align}
and calculate the correlation functions
\begin{equation} \label{3yCorrelation}
  \left\langle y(p_1) \cdots y(p_k) \right\rangle = \frac{(-1)^k}{\prod_{i=1}^k \sin^2 p_i}
  \left\langle \prod_{i=1}^k \sum_{n=1}^N \frac{1}{q_i + z_n} \right\rangle
\end{equation}
from which the correlation functions \eqref{2kPointCorrelationDef} can
always be reconstructed. Expanding the $k$-fold product over the
$w(p_i) = y(p_i) + N q_i$, we arrive at
\begin{equation} \label{3GeneralizedBinomial}
C_k\left( p_1, \ldots, p_k \right) = \sum_{i=0}^k \sum_{\omega \in \mathbb{S}_k} \frac{ N^{k-i}}{i!(k-i)!}
  \left(\prod_{l=1}^{k-i} q_{\omega(l)}\right) \left\langle \prod_{l=k-i+1}^k y\left(p_{\omega(l)}\right) \right\rangle .
\end{equation}
The second sum runs over all elements $\omega(l)$ in the permutation
group $\mathbb{S}_k$ of $k$ objects. It enters the formula, because the correlation functions \eqref{3yCorrelation} appear in all orders $i$ up to $k$, comprising different subsets of $\{p_1, \ldots, p_k\}$ with
cardinality $i$. Thus, the $k$-point correlation function $C_k$ can be
determined from all lower order correlation functions
\eqref{3yCorrelation}.

Performing the product, the average \eqref{3yCorrelation} becomes a
complicated sum of terms. In some of them, only one of the eigenvalues
$z_n$ appears, these are the disconnected parts of the average to be
performed. All other terms contain at least two different eigenvalues
and may thus be referred to as connected. However, in Sec.~\ref{sec3}
we will rewrite the ensemble average in Eq.~\eqref{3yCorrelation} in
such a way that all terms can be obtained from the average of the $N$-
point completely connected average 
\begin{equation} \label{NPointConnectedAverage}
\left\langle\prod_{n=1}^N\frac{1}{q_n+z_n}\right\rangle,
\end{equation}
which is, due to its very definition as an average, invariant under
all permutations of the $N$ arguments $q_n , \ n=1,\ldots,N$. Our correlation functions, however,
only depend on $k$ of those arguments $q_n , \ i=1,\ldots,k$ where we assume
$k\le N$. We find the proper $k$-point connected average by taking the limit
\begin{equation} \label{3kPointDisconnectedPartLimit}
  \left\langle\prod_{n=1}^k\frac{1}{q_n+z_n}\right\rangle =
  \lim_{q_{k+1},\ldots, q_N\to\infty}\left(\prod_{m=k+1}^Nq_m \right)\left\langle\prod_{n=1}^N\frac{1}{q_n+z_n}\right\rangle,
\end{equation}
over the $N-k$ excess variables $q_i , \ i=(k+1),\ldots,N$. For this
$N$-point connected average \eqref{NPointConnectedAverage} we derive
in Sec.~\ref{sec3} the result
\begin{equation} \label{3NPointDisconnectedPartNdet}
  \left\langle\prod_{n=1}^N\frac{1}{q_n+z_n}\right\rangle =
  \frac{1}{c_N\pi^N}\sum_{\omega \in \mathbb{S}_N}\det\left[L_{n m \omega(n)}(q_{\omega(n)})\right]_{n,m=1,\ldots,N}
\end{equation}
with the function
\begin{equation}\label{FunctionL}
L_{nml}(q_l) = \frac{(-1)^{m-n} \pi}{q_l^{m-n+1}} B(m,N-m+1)
\begin{cases}
u_m(N,q_l^2) \qquad &m\geq n
\\
-v_m(N,q_l^2) \qquad &m<n
\end{cases}.
\end{equation}
The functions $u_m(N,q_l^2)$ and $v_m(N,q_l^2)$ are given by
\begin{align} \label{3ReducedBetaFunctions}
u_m(N,q_l^2) &= \frac{2}{B (m,N-m+1)} \int_0^{q_l} d\rho \frac{\rho^{2m-1}}{(1+\rho^2)^{N+1}} \nonumber
\\
v_m(N,q_l^2) &= \frac{2}{B (m,N-m+1)} \int_{q_l}^\infty d\rho \frac{\rho^{2m-1}}{(1+\rho^2)^{N+1}}
\end{align}
and may be viewed as normalized incomplete Beta functions with the property
\begin{equation} \label{3ReducedBetaFunctionsP}
u_m(N,q_l^2) + v_m(N,q_l^2) = 1 .
\end{equation}
We will come across these functions also in the distribution of the winding number to be discussed in Sec.~\ref{sec22}.
Taking the limit \eqref{3kPointDisconnectedPartLimit}, the result \eqref{3NPointDisconnectedPartNdet} yields
the $k$-point connected average,
\begin{equation} \label{4DisconnectedPartkDet}
  \left\langle\prod_{n=1}^k\frac{1}{q_n+z_n}\right\rangle =
  \frac{1}{c_N \pi^k} \sum_{\omega \in \mathbb{S}_N} \left( \prod_{l=k+1}^N B(\omega(l), N-\omega(l)+1) \right)
     \det \left[L_{\omega(m)\omega(n)n}(q_n)\right]_{n,m=1,\ldots,k}
\end{equation}
which is a $k\times k$ determinant, as derived in Sec.~\ref{sec3}.

From the general formulae~\eqref{3GeneralizedBinomial} and \eqref{4DisconnectedPartkDet}, we obtain in Sec.~\ref{sec3}
for the first two correlation functions
\begin{align} \label{3kPointCorrelationResults}
C_1 (p_1) &= 0
\\
C_2 (p_1,p_2) &= -\frac{1-\cos^{2N} \left(p_1 - p_2\right)}{1-\cos^2 \left(p_1 - p_2\right)}.	\nonumber
\end{align}
We notice that the two-point function depends only on the distance between
the points $p_1$ and $p_2$ on the parameter manifold, which is a
consequence of the translation invariance of our random matrix field
\eqref{2RMF}. It turns out that for all $k$ one of the parameters can be set to
zero (or any other arbitrary point) without losing any information.

\subsection{Universality Aspects and Unfolding of the Two-Point Function}
\label{sec21a}

The power of Random Matrix Theory lies in the universality of its
statistical predictions.  For matrix dimensions tending to infinity, the spectral
correlations measured on the local scale of the mean level spacing $\Delta$
coincides for all probability densities of the random matrices that do
not have scales competing with the mean level
spacing~\cite{GuhrRMTReview, MehtaBook}. The required rescaling
procedure is referred to as unfolding. Furthermore, a similar
universality is valid for the parametric
correlations \cite{BeenakkerRejaei1994}.
Inspired by this, we search for universal regimes in our correlation functions.
To this end, we rescale the parameters with a positive power of $N$ according to
\begin{equation} \label{3Rescaling}
\psi_i = N^\alpha p_i . 
\end{equation}
We consider positive powers because we want to zoom into the
parametric dependence to observe it on a proper local
scale in the limit $N \rightarrow \infty$. Naturally, all physics
systems that we want to compare with our random matrix theory should
be considered on the same scale.

\begin{figure}\label{fig1}
\centering
\includegraphics[width=.9\linewidth]{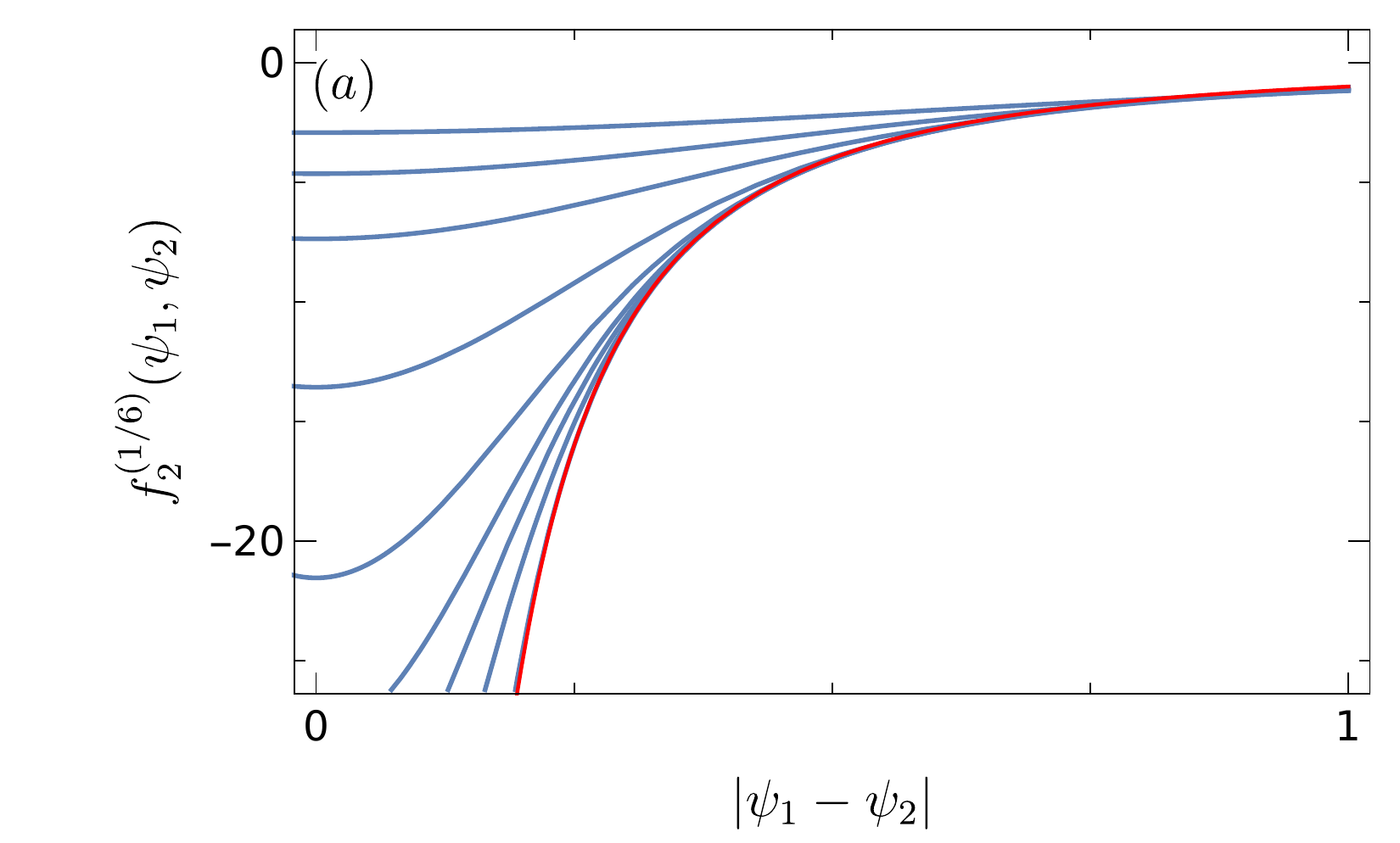}
\\
\includegraphics[width=.9\linewidth]{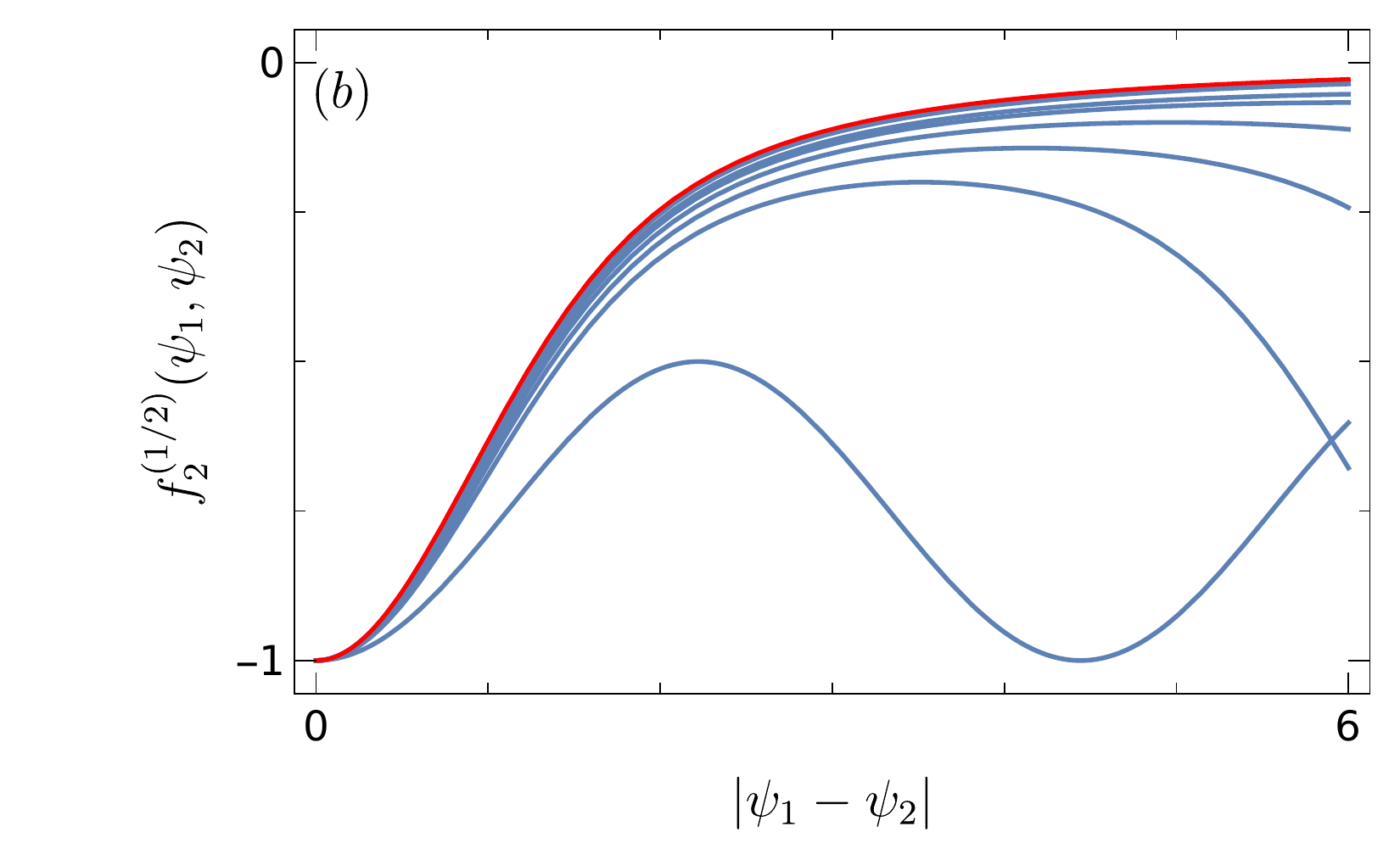}
\caption{Unfolded two-point function after the rescaling \eqref{3Rescaling} for different values of $N$ (blue). In (a) we used $N = 5, 10, 20, 50, 100, 150, 200, 300, 1000$ and $\alpha = 1/6$, in (b) $N = 2, 5, 7, 10, 15, 20, 50, 100$ and $\alpha = 1/2$. For comparison the limit \eqref{3UnfoldingLimit} (red).}
\end{figure}

We turn to the two-point function
\eqref{3kPointCorrelationResults}. In the limit of large $N$ the
rescaled arguments $\psi_i/N^\alpha$ become small, allowing us to
expand the cosines. We find
\begin{equation} \label{3UnfoldingLimit}
  \lim_{N\rightarrow \infty} C_2\left(\frac{\psi_1}{N^\alpha},\frac{\psi_2}{N^\alpha}\right)
             \frac{d\psi_1}{N^\alpha} \frac{d\psi_2}{N^\alpha} = f_2^{(\alpha)}(\psi_1,\psi_2) d\psi_1 d\psi_2
\end{equation}
with the function
\begin{equation}\label{3UnfoldingLimitB}
f_2^{(\alpha)}(\psi_1,\psi_2) = \begin{cases}
\displaystyle -\frac{1}{\left(\psi_1 - \psi_2\right)^2} \qquad &\alpha < \frac{1}{2}
\\
\displaystyle -\frac{1-\exp(-( \psi_1 - \psi_2)^2)}{\left( \psi_1 - \psi_2\right)^2} \qquad &\alpha = \frac{1}{2}
\\
\displaystyle 0 \qquad &\alpha > \frac{1}{2}
\end{cases}.
\end{equation}
The case $p_1=p_2$ or $\psi_1=\psi_2$, respectively, is subject to
interpretation. As obvious from Eq.~
\eqref{3kPointCorrelationResults}, we have $C_2(p_1,p_1)=-1$. Hence,
we must assume that the arguments are not equal, $\psi_1\neq\psi_2$,
when taking the limit for arbitrary $\alpha$. We observe different
regimes in the result \eqref{3UnfoldingLimitB}. The regime with
$\alpha = 1/2$ amounts to an unfolding with the inverse of the mean
level spacing $\Delta = \pi/\sqrt{2N}$ of the Gaussian unitary
ensemble (GUE). This is in accordance with the works on parametric
level correlations \cite{SimonsAltshuler1993A, SimonsAltshuler1993B},
where a universal statistic was found after rescaling the parameter
space with the root of the mean square velocity $\sqrt{\left\langle
  \left(\partial E_m(p_i)/\partial p_i\right)^2
  \right\rangle/\Delta^2}$ of the energy $E_m=E_m(p_i)$. In the
present case $\sqrt{\left\langle \left(\partial E_m(p)/\partial
  p_i\right)^2 \right\rangle} = 1$ can be found using the methods of
Ref.~\cite{BeenakkerRejaei1994}.  In Fig.~\ref{fig1} we display our result for two choices of $\alpha$ and various values of $N$. As seen,
the unfolded two-point function approaches the limit
\eqref{3UnfoldingLimitB} when $N$ increases. We conjecture that the
function $f_2^{(\alpha)}(\psi_1,\psi_2)$ is universal.

\subsection{Winding Number Distribution}
\label{sec22}

For the discussion to follow, it is useful to cast the random matrix
field \eqref{2RMF} into an equivalent, but different form.  Introducing $s =
e^{ip}$ as complex variable on the unit circle, we have
\begin{equation} \label{3RMF}
K(s) = \frac{s}{2}( K_1 - i K_2) + \frac{1}{2s} ( K_1 + i K_2).
\end{equation}
For the determinant we have
\begin{align}	\label{3RMFDet}
\det K(s) &= \frac{1}{(2s)^N} \det (K_1 + iK_2 + s^2(K_1 - i K_2))
\\
&= \frac{\det(K_1 - iK_2)}{(2s)^N} \prod_{n=1}^N \left(s^2 + z_n'\right),	\nonumber
\end{align}
where the $z_n'$ are the solution of the generalized eigenvalue problem
\begin{equation} \label{3GeneralizedEigenvalueProblem}
( K_1 + i K_2 ) \vc{v}_n = z_n' \left( K_1 - i K_2 \right) \vc{v}_n
\end{equation}
with eigenvectors $\vc{v}_n$. The matrices $K_1 \pm i K_2 $ are again
Ginibre matrices, implying that the probability distribution of the
$z_n'$ is the one of the spherical ensemble
\eqref{3SphericalEnsembleDistribution}. In the sequel, we thus always
write $z_n$. The winding number in terms of Eq.~\eqref{3RMF} is
\begin{equation}
W = \frac{1}{2\pi i}\oint_{\abs{s}=1}ds \frac{1}{\det K(s)} \frac{d}{ds} \det K(s) 
  = \frac{1}{2\pi i}\oint_{\abs{s}=1}ds \frac{d}{ds} \ln\det K(s) .
\end{equation}
Obviously, $\det(K_1-iK_2)$ drops out in the integrand. The
contour integral yields the difference of zeros and poles of $\det
K(s)$ inside the unit circle. From Eq.~\eqref{3RMFDet} we infer that
it has a pole of order $N$ at zero and that its zeros come in pairs,
making their number even. Let $m$ be the number of solutions of
Eq.~\eqref{3GeneralizedEigenvalueProblem} that lie inside the unit
circle, then
\begin{equation}\label{Wm}
W = 2m - N
\end{equation}
is the winding number.  The number $m$ takes values from $0$ to $N$,
thus the winding number lies between $-N$ and $N$. The probability that
$m$ eigenvalues are inside the unit circle and the remaining ones outside is
\begin{equation} \label{3qProbabilityD}
r(m) = \int\limits_{\abs{z_1} < 1} d[z_1] \cdots \int\limits_{\abs{z_m} < 1} d[z_m]
        \int\limits_{\abs{z_{m+1}} > 1} d[z_{m+1}] \cdots \int\limits_{\abs{z_N} > 1} d[z_N] \, G(z) .
\end{equation}
In Sec.~\ref{sec3} we show that
\begin{equation} \label{3qProbabilityR}
r(m) = \frac{1}{N!} \sum_{\omega \in \mathbb{S}_N} \left(\prod_{i=1}^m u_{\omega(i)}(N,1)\right)
                 \left(\prod_{i=m+1}^N v_{\omega(i)}(N,1)\right),  
\end{equation}
where the expressions $u_i(N,1)$ and $v_i(N,1)$ follow from the
functions Eq.~\eqref{3ReducedBetaFunctions}. Taking into account the
permutation invariance of the eigenvalues inside, respectively outside, the unit circle
and using Eq.~\eqref{Wm} we find the discrete
probability distribution
\begin{equation} \label{3WindingNumberDistribution}
P(W) = r\left(\frac{W+N}{2}\right) {N \choose (W+N)/2}
\end{equation}
on the integers $W$ between $-N$ and $N$ as the winding number
distribution for arbitrary, finite matrix dimension $N$.

Let us now turn to the moments \eqref{2WindingNumberMomentsDef} of
this distribution. Since the one-point function
\eqref{3kPointCorrelationResults} vanishes, the mean winding number is
zero
\begin{equation} \label{3WMean}
\left\langle W \right\rangle = 0.
\end{equation}
To arrive at a closed form for $k=2$ we calculate, instead of directly
applying the definition \eqref{2WindingNumberMomentsDef}, the
difference in the winding number variance of systems with
$(N+1)\times(N+1)$ and $N\times N$ dimensional chiral subblocks. The
second moment is given by
\begin{equation}
  \left\langle W^2 \right\rangle\Big|_N = -\frac{1}{4\pi^2} \int_0^{2\pi} dp_1 dp_2\ C_2(p_1,p_2) 
  = \frac{1}{2\pi} \int_0^{2\pi} d\varphi \frac{1-\cos^{2N} \varphi}{1-\cos^2 \varphi}, 
\end{equation}
where we indicate the $N$ dependence. For the difference we find
\begin{align}
  \left\langle W^2 \right\rangle\Big|_{(N+1)} - \left\langle W^2 \right\rangle\Big|_N
  &= \frac{1}{2\pi} \int_0^{2\pi} d\varphi \cos^{2N} \varphi \\
  &= \frac{(2N-1)!!}{(2N)!!} = \frac{(2N+1)!!}{(2N)!!} - \frac{(2N-1)!!}{(2N-2)!!},	\nonumber
\end{align}
and with $\left\langle W^2 \right\rangle|_1 = 1$ we obtain
\begin{equation} \label{3WVariance}
\left\langle W^2 \right\rangle = \frac{(2N-1)!!}{(2N-2)!!} \simeq 2 \sqrt{\frac{N}{\pi}}.
\end{equation}
The last expression holds for large $N$. Hence, the second moment
grows with $\sqrt{N}$, not with $N$. The results \eqref{3WMean} and
\eqref{3WVariance} suggests to look at the distribution of $P(W)$ as a
function of $W^2/\sqrt{N}$ for large $N$. Numerically, we find that it is well
described by
\begin{equation}
\frac{P(W)}{P(0)} = \exp \left(-\frac{1}{4} \sqrt{\frac{\pi}{N}} W^2\right),
\end{equation}
i.e.~by a Gaussian distribution. 
\section{Derivations}
\label{sec3}

In Sec.~\ref{sec31} we reformulate the quantity to be ensemble averaged
in the $k$-point correlation function \eqref{3yCorrelation}. We
calculate the $N$-point and the $k$-point connected ensemble averages
in Secs.~\ref{sec32} and~\ref{sec32a}, respectively. The explicit
expressions for the one and two-point functions are worked out in
Sec.~\ref{sec33}. In Sec.~\ref{sec34} we compute the probability
\eqref{3qProbabilityD} appearing in the discrete winding number
distribution \eqref{3WindingNumberDistribution}.

\subsection{Reformulation of the Key Expression to be Ensemble Averaged}
\label{sec31}

\begin{figure}
\centering
  \includegraphics[width=.6\linewidth]{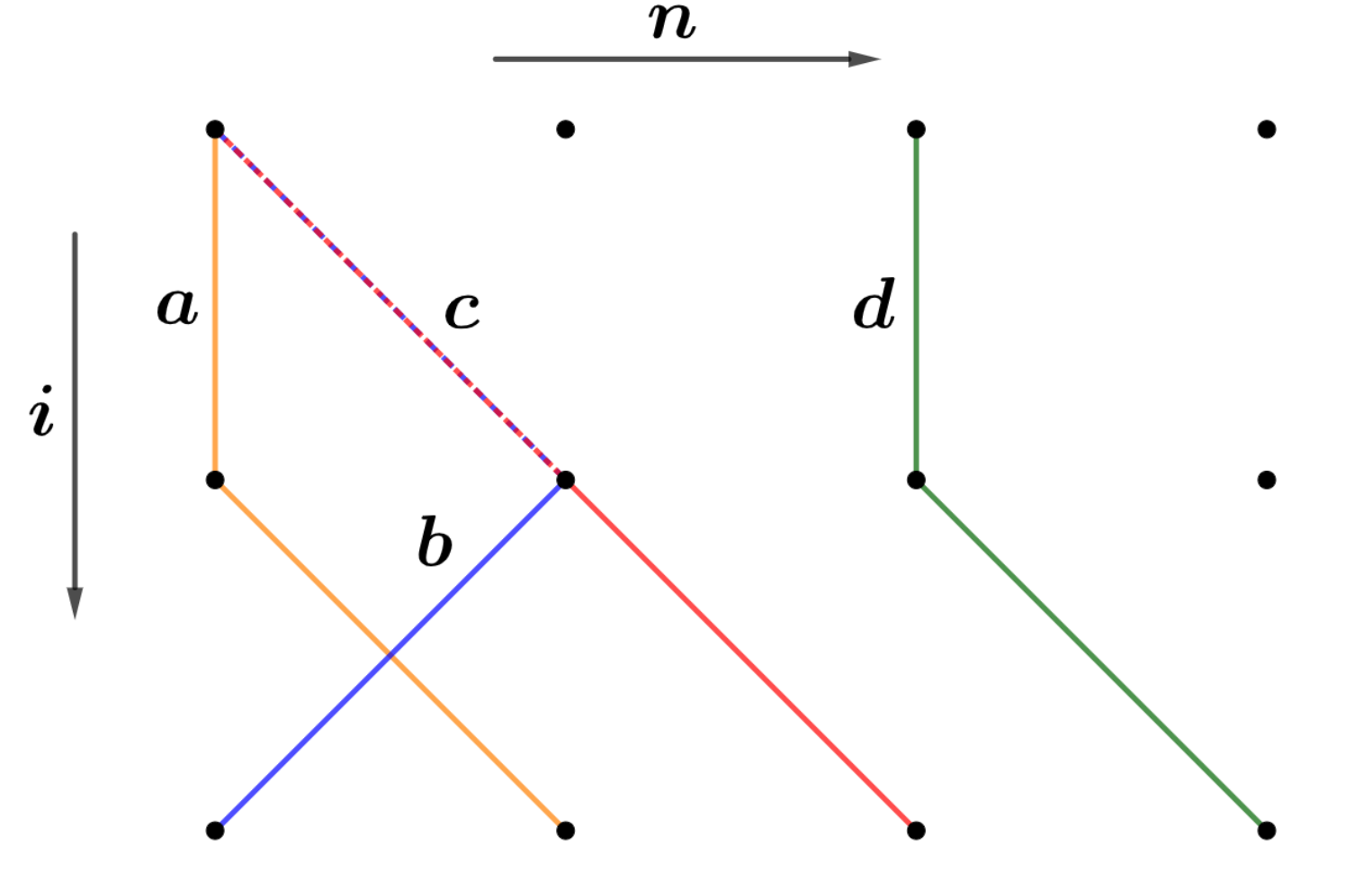}
  \caption{Examples for paths on a $3\times 4$ lattice. Paths $\vc{a}$
    and $\vc{b}$ both have the multiindex $(2,1,0,0)$ and are thus
    equivalent up to permutations in $i$. Path $\vc{c} = (1,1,1,0)$ is
    the completely connected path along the angle bisector of the lattice.
    Path $\vc{d} = (0,0,2,1)$ is equal to path $\vc{b}$ as the
    integration variables $z_n$ are permutation
    invariant.}
  \label{4ConnectedDisconnectedPaths}
\end{figure}

To perform the calculation of the correlation function \eqref{3yCorrelation}, it is helpful
to rewrite the expression to be ensemble averaged, namely
\begin{equation}\label{Average}
\left\langle \prod_{i=1}^k \sum_{n=1}^N \frac{1}{q_i + z_n} \right\rangle,
\end{equation}
by pulling out, pictorially speaking, the sums from the angular brackets,
i.e.~to cast the average \eqref{Average} into a sum of terms
containing only products to be averaged. This requires some work. We
use the permutation invariance of the distribution
\eqref{3SphericalEnsembleDistribution} and think of the product of sums as
a $k \times N$ lattice. Let the rows be labelled by $i = 1,\ldots, k$
and the columns by $n = 1,\ldots, N$. As depicted in
Fig.~\ref{4ConnectedDisconnectedPaths} for some examples, each term in the product is a
path through the lattice, obeying the following rules:
\begin{itemize}
\item Each row is visited once and only once. This amounts to each
  $q_i$ appearing only once in each of the terms.
\item Two paths are considered equal if they visit the same lattice
  points, irrespective of the order. The points on the lattice are
  coupled via multiplication, which is commutative.
\end{itemize} 
To each path we assign a multiindex $\vc{l} = (l_1,\ldots,l_N) \in
\mathbb{N}_0^N$ of length $\abs{\vc{l}} = \sum_{n=1}^N l_n = k$. It
describes how many times $l_n$ the path has visited the $n$--th column
and therefore how many factors including $z_n$ appear in the
associated term. However, this mapping is not unique. There are in
total
\begin{equation}
{k \choose \vc{l}} = \frac{k!}{l_1! \cdots l_N!}
\end{equation}
paths sharing the same $\vc{l}$. In the matrix average these terms are
equal up to permutations in the $q_i$. We take care of this by setting
$q_i \rightarrow q_{\omega (i)}$ and summing over all permutations
\begin{equation}
  \left\langle \prod_{i=1}^k \sum_{n=1}^N \frac{1}{q_i + z_n} \right\rangle =
  \frac{1}{k!} \sum_{\omega \in \mathbb{S}_k} \sum_{\abs{\vc{l}}=k}
   \left\langle \prod_{i=1}^k \frac{1}{q_{\omega(i)} + z_{g_{\vc{l}}(i)}} \right\rangle.
\end{equation}
Here, we introduce the step function
\begin{equation}
g_{\vc{l}}(i) = 1 + \sum_{n=1}^N \Theta \left( i - \sum_{j=1}^n l_j \right) , \quad \text{with} \quad \Theta (0) = 0,
\end{equation}
employing the Heaviside unit step function $\Theta$, to select the correct
variables $z_n$ for the integration of the corresponding
product. We distinguish between different types of paths. In the disconnected paths only one $z_n$ appears, which amounts to $l_n = k$ for one $n$ and $l_n = 0$ for all other $n$. We refer to all other paths as connected. Out of the connected paths the ones with $l_n \in \{0,1\}$, where each $z_n$ may appear only once, stand out. To these paths we refer as completely connected and their contributions may be evaluated via Eq. \eqref{3kPointDisconnectedPartLimit}.

Next we consider the permutation invariance of the $z_n$. Let
$h_{\vc{l}}(i)$ be the function that tallies up the number of integers
$i$ appearing in $\vc{l}$. There are
\begin{equation}
\frac{N!}{\prod_{i=1}^k h_{\vc{l}}(i)!}
\end{equation}
possible ways to permute the $z_n$ without changing the ensemble
average. We choose the ordered multiindex $\vc{l}$ with $l_1 \leq
\ldots \leq l_N$ as a representative for all of these paths. On
the $k \times N$ lattice, this amounts to paths below the angle
bisector. We thus finally arrive at
\begin{equation} \label{4GeneralizedLaplace}
  \left\langle \prod_{i=1}^k \sum_{n=1}^N \frac{1}{q_i + z_n} \right\rangle =
  \frac{1}{k!} \sum_{\omega \in \mathbb{S}_k} \sum\limits_{\substack{l_1 \leq \ldots \leq l_N \\
      \abs{\vc{l}} = k}} {k\choose \vc{l}}
      \frac{N!}{\prod_{i=1}^k h_{\vc{l}}(i)!} \left\langle \prod_{i=1}^k \frac{1}{q_{\omega(i)} + z_{g_{\vc{l}}(i)}} \right\rangle.
\end{equation}
Indeed, this is a sum over ensemble averages of products only. Generally, any $z_n$ may appear $l$ times. To handle this, we use
the partial fraction expansion
\begin{equation} \label{4PartialFractionExpansion}
\prod_{i=1}^l \frac{1}{q_i + z_n} = \sum_{i=1}^l \frac{1}{\prod_{j\neq i} (q_j - q_i)} \frac{1}{q_i + z_n},
\end{equation}
which reduces the corresponding averages to a sum of completely connected averages. Thus, the
resulting expression can again be treated with Eq.~\eqref{3kPointDisconnectedPartLimit}.

\subsection{Calculation of the $N$-Point Connected Ensemble Average}
\label{sec32}

As already pointed out in Sec.~\ref{sec21}, all connected $k$-point ensemble averages
can be, via proper limits, obtained from the connected $N$-point average
\begin{equation} \label{4NPointDisconnectedPart}
\left\langle\prod_{n=1}^N\frac{1}{q_n+z_n}\right\rangle=\int d[z] \, G(z)\prod_{n=1}^N\frac{1}{q_n+z_n},
\end{equation}
where $G(z)$ is the joint probability density
\eqref{3SphericalEnsembleDistribution} of the spherical ensemble. We use
\begin{equation} \label{VandermoneE}
|\Delta_N(z)|^2 = \Delta_N(z) \Delta_N^*(z) = \Delta_N(z) \Delta_N(z^*)
\end{equation}
and expand the Vandermonde determinant $\Delta_N(z)$ in the Laplace form. This yields
\begin{align}
  \left\langle\prod_{n=1}^N\frac{1}{q_n+z_n}\right\rangle &=
  \frac{1}{c_N\pi^N} \sum_{\omega \in \mathbb{S}_N} \text{sgn}\, \omega
  \int d[z]\Delta_N(z^*)\prod_{n=1}^N \frac{z_n^{\omega(n)-1}}{\left(1+\abs{z_n}^2\right)^{N+1}\left(q_n+z_n\right)} \nonumber \\
  &= \frac{1}{c_N\pi^N} \sum_{\omega \in \mathbb{S}_N} \int d[z] \Delta_N(z^*)
    \prod_{n=1}^N \frac{z_n^{n-1}}{\left(1+\abs{z_n}^2\right)^{N+1}\left(q_{\omega(n)}+z_n\right)}
\end{align} 
where the second equation follows from renaming the integration
variables $z_n \rightarrow z_{\omega (n)}$ for each permutation
$\omega \in \mathbb{S}_N$.  The sign $\text{sgn}\, \omega$ of the
permutation $\omega$ is canceled by the same sign appearing in
$\Delta_N(z^*)$ when changing the integration variables. Inserting the
remaining Vandermonde determinant and integrating row by row we obtain
Eq.~\eqref{3NPointDisconnectedPartNdet} with the function
\begin{align} \label{4LFunction}
L_{nml}(q) &=\int d[z_n]\frac{\left(z_n^*\right)^{m-1}z_n^{n-1}}{\left(1+\abs{z_n}^2\right)^{N+1}\left(q_l + z_n\right)} \nonumber \\
&= \int_0^\infty d\rho_n \frac{\rho_n^{m+n-1}}{(1 + \rho_n^2)^{N+1}} \int_0^{2\pi} d\vartheta_n \frac{e^{i(n-m)\vartheta_n}}{q_l + \rho_n e^{i\vartheta_n}},
\end{align}
where we employ polar coordinates $z_n=\rho_ne^{i\vartheta_n}$ in the
second equation. The angular integral yields, by virtue of the residue
theorem,
\begin{equation}
\int_0^{2\pi} d\vartheta_n \frac{e^{i(n-m)\vartheta_n}}{q_l + \rho_n e^{i\vartheta_n}} =
\begin{cases} \displaystyle
\frac{2\pi}{q_l} \left( -\frac{\rho_n}{q_l} \right)^{m-n} \qquad & m\geq n, \rho_n < q_l
\\ \displaystyle
\frac{2\pi}{\rho_n} \left( -\frac{\rho_n}{q_l} \right)^{m-n+1} \qquad & m<n, \rho_n > q_l
\\ \displaystyle
0 \qquad &\text{else}
\end{cases}.
\end{equation}
Thus we arrive at Eq.~\eqref{FunctionL}.

\subsection{Reduction to the $k$-Point Connected Ensemble Average}
\label{sec32a}

To take the limit \eqref{3kPointDisconnectedPartLimit} we  need as an intermediate result
a proper limit involving the function $L_{nml}(q_l)$. As the limit $q_l\to\infty$ of the incomplete Beta functions
\eqref{3ReducedBetaFunctions} gives either unity or zero,
the total limit is only non--vanishing if $m=n$,
\begin{equation} \label{4LLimit}
\lim_{q_l \rightarrow \infty} q_l L_{nml}(q_l) = 
\begin{cases}
\pi B(m,N-m+1) \qquad &m=n
\\
0 \qquad &m\neq n
\end{cases}.
\end{equation}
We apply this result to reduce the $k$-point connected average, which
is, according to Eqs.~\eqref{3kPointDisconnectedPartLimit} and
\eqref{3NPointDisconnectedPartNdet}, a limit of an $N\times N$
determinant.  The limit makes all elements in the $\omega^{-1}(n)$--th
row vanish except the diagonal element, which is $\pi
B(\omega^{-1}(n),N-\omega^{-1}(n)+1)$. We expand the determinant in
these elements
\begin{align}
  \left\langle\prod_{n=1}^k\frac{1}{q_n+z_n}\right\rangle &=
   \frac{1}{c_N \pi^k} \sum_{\omega \in \mathbb{S}_N} \left( \prod_{l=k+1}^N B(\omega^{-1}(l),N-\omega^{-1}(l)+1) \right) \nonumber \\
   & \qquad\qquad\qquad \det\left[L_{n m \omega(n)}(q_{\omega(n)})\right]_{n,m=1,\ldots,N}^{n,m\neq \omega^{-1}(l), l=k+1,\ldots,N}.
\end{align}
Interchanging row $n$ with row $\omega^{-1}(n)$ and column $m$ with column $\omega^{-1}(m)$
yields for the right hand side
\begin{align}
& \frac{1}{c_N \pi^k} \sum_{\omega \in \mathbb{S}_N} \left( \prod_{l=k+1}^N B(\omega^{-1}(l),N-\omega^{-1}(l)+1) \right) 
\det\left[L_{\omega^{-1}(n) \omega^{-1}(m) n}(q_n)\right]_{n,m=1,\ldots,k}	\nonumber
\\
&\qquad\qquad  = \frac{1}{c_N \pi^k} \sum_{\omega \in \mathbb{S}_N} \left( \prod_{l=k+1}^N B(\omega(l),N-\omega(l)+1) \right) 
\det\left[L_{\omega(n) \omega(m) n}(q_n)\right]_{n,m=1,\ldots,k}.
\end{align}
We also used that the order in the sum over the permutations $\omega$
is invariant due to the group property of $\mathbb{S}_N$.  Thus, we
arrive at the result \eqref{4DisconnectedPartkDet}.

\subsection{Explicit Expressions for the One and Two-Point Correlation Functions}
\label{sec33}

For $k=1$ there are no connected terms. According to
Eq.~\eqref{3GeneralizedBinomial} and \eqref{4GeneralizedLaplace} the
one-point function is given by
\begin{equation}
C_1(p_1) = \left\langle y(p_1) \right\rangle + Nq_1
= -\frac{N}{\sin^2 p_1} \left\langle \frac{1}{q_1 + z_1} \right\rangle + Nq_1.
\end{equation}
The average follows from Eq.~\eqref{4DisconnectedPartkDet},
\begin{equation} \label{4OnePointAverage}
\left\langle \frac{1}{q_1+z_1} \right\rangle = \frac{1}{Nq_1} \sum_{n=1}^N u_n(N,q_1^2).
\end{equation}
The incomplete Beta functions \eqref{3ReducedBetaFunctions} may be
rewritten using integration by parts, we find 
\begin{equation}
v_m(N,q_1^2) = \sum_{l=0}^{m-1} {N-1-l \choose m-1-l}\frac{(q_1^2)^{m-l-1}}{(1+q_1^2)^{N-l}}.
\end{equation}
Using the property \eqref{3ReducedBetaFunctionsP} the sum in
Eq.~\eqref{4OnePointAverage} can be evaluated
by means of the binomial theorem, implying
\begin{equation} \label{41PointDisonnectedPart}
\left\langle \frac{1}{q_1+z_1} \right\rangle = \frac{1}{q_1} - \frac{1}{q_1(1+q_1^2)} = \sin p_1 \cos p_1.
\end{equation}
In the last step we reinserted $q_1 = \cot p_1$. Altogether we arrive at the first 
of the results \eqref{3kPointCorrelationResults}.

For $k=2$ we apply formulae \eqref{3GeneralizedBinomial} and
\eqref{4GeneralizedLaplace} and use the vanishing of the one-point
function
\begin{align}
C_2(p_1,p_2) &= \left\langle y(p_1) y(p_2) \right\rangle - N^2 q_1 q_2 \nonumber \\
&= \frac{1}{\sin^2 p_1 \sin^2 p_2} \left\langle \prod_{i=1}^2 \sum_{n=1}^N \frac{1}{q_i + z_n} \right\rangle - N^2 q_1 q_2.	
\end{align}
With Eq.~\eqref{4GeneralizedLaplace} we find
\begin{equation}
\left\langle \prod_{i=1}^2 \sum_{n=1}^N \frac{1}{q_i + z_n} \right\rangle = N \left\langle \frac{1}{q_1+z_1} \frac{1}{q_2+z_1} \right\rangle + N(N-1) \left\langle \frac{1}{q_1+z_1} \frac{1}{q_2+z_2} \right\rangle.
\end{equation}
The connected average is given by \eqref{4DisconnectedPartkDet} and reads
\begin{align}
&\left\langle \frac{1}{q_1+z_1} \frac{1}{q_2+z_2} \right\rangle = \frac{1}{N(N-1)}  \frac{1}{q_1 q_2} 
\left( \sum\limits_{\substack{n,m=1 \\ n\neq m}}^N u_n(N,q_1^2) u_m(N,q_2^2) \right. \nonumber
\\
&\qquad\qquad + \left. \sum\limits_{\substack{n,m=1 \\ n > m}}^N \left(\frac{q_2}{q_1}\right)^{n-m} u_n(N,q_1^2) v_m(N,q_2^2)
         + \left(\frac{q_1}{q_2}\right)^{n-m} v_n(N,q_1^2) u_m(N,q_2^2) \right)	
\end{align}
This expression is readily simplified by using the translation
invariance on the parameter manifold. We set $p_2 = \pi/2$ which amounts
to $q_2 = 0$ and find
\begin{align}
  \left\langle \frac{1}{q_1+z_1} \frac{1}{z_2} \right\rangle &= \frac{1}{N(N-1)} \frac{1}{q_1^2} \sum_{n=2}^N u_n(N,q_1^2) \nonumber\\
  & = \frac{1}{(N-1)q_1^2} \left( 1-\frac{1}{1+q_1^2} + \frac{1}{N(1+q_1^2)^N} -\frac{1}{N}\right).
\end{align}
For the disconnected average we employ the partial fraction expansion
\eqref{4PartialFractionExpansion} and Eq.~\eqref{41PointDisonnectedPart},
\begin{equation}
  \left\langle \frac{1}{q_1+z_1} \frac{1}{z_1} \right\rangle
  = -\frac{1}{q_1}\left( \left\langle \frac{1}{q_1+z_1} \right\rangle - \left\langle \frac{1}{z_1} \right\rangle \right)
  = -\frac{1}{1+q_1^2}.
\end{equation}
Reinserting $q_1 = \cot p_1$ yields
\begin{equation}
C_2\left(p_1,\frac{\pi}{2}\right) = -\frac{1-\cos^{2N} p_1}{1-\cos^2 p_1} 
\end{equation}
or, equivalently, the second of the results \eqref{3kPointCorrelationResults}.

\subsection{Calculation of the Probability $r(m)$}
\label{sec34}

For the discrete winding number distribution
\eqref{3WindingNumberDistribution} we need to compute the
probability \eqref{3qProbabilityD}. The calculation is similar to the
one in Sec.~\ref{sec32}.  Inserting the probability density
\eqref{3SphericalEnsembleDistribution}, treating the Vandermonde
determinants as in Sec.~\ref{sec32} and renaming the integration
variables $z_n \rightarrow z_{\omega(n)}$, we have
\begin{align}
r(m) &= \frac{1}{c_N \pi^N} \sum_{\omega \in \mathbb{S}_N} \text{sgn}\, \omega
  \int\limits_{\abs{z_{\omega(1)}} < 1} d[z_1] \cdots \int\limits_{\abs{z_{\omega(m)}} < 1} d[z_m] \nonumber \\
  & \qquad\qquad \int\limits_{\abs{z_{\omega(m+1)}} > 1} d[z_{m+1}] \cdots \int\limits_{\abs{z_{\omega(N)}} > 1} d[z_N]
           \, \prod_{n=1}^N \frac{z_n^{n-1} (z_n^*)^{\omega(n)-1}}{(1 + \abs{z_n}^2)^{N+1}}.
\end{align}
With polar coordinates $z_n = \rho_n e^{i\vartheta_n}$ we find Kronecker deltas for the angular integrals,
\begin{equation}
\int_0^{2\pi} d\vartheta_n e^{i(n-\omega(n))\vartheta_n} = 2\pi \delta_{n\omega(n)}.
\end{equation}
Thus, only the unit permutation $\omega = \unit$ contributes. The
radial integrals are given by the functions
\eqref{3ReducedBetaFunctions} for $q_l=1$.  Altogether we arrive at
formula \eqref{3qProbabilityR}.

\section{Conclusions}
\label{sec4}

We studied the winding number in a model of parameter dependent chiral
random matrices. This seems to be the first time that statistical
topology for a chiral symmetry class has been studied in such a
schematic model. Apart form the conceptual importance, the winding
number has concrete physics interpretations, for example, as the
topological index belonging to gapped phases of fermions.  We found
that the joint probability density of the complex eigenvalues in our
model coincides with that of the spherical ensemble which is
known in the literature. We used it to address the new questions of
statistical topology, we analytically calculated the discrete
probability distribution of the winding numbers, as well as the
parametric correlations functions of the winding number density.  We
derived a closed formula for the former and arrived for the latter at
explicit determinant expressions for certain correlation functions of
arbitrary order which allow for a construction of the
winding number density correlations functions.  We constructed the one
and two-point functions.  All our results involve incomplete Beta
functions which are fairly simple.

As Random Matrix Theory is widely known to provide universal results
for spectral statistics and certain parametric statistics, we are
confident that our results hold universal information as well. To
reveal it, we carried out an unfolding procedure similar to the one in
the above contexts. Remarkably, we found different scaling limits. We
expect our results for the unfolded two-point correlation function to
be universal.

Our results, namely the implied universality of the correlation function and the Gaussian distribution of the topological index, are analogous to the ones obtained numerically in the case of the adiabatic curvature and the Chern number \cite{GatWilkinson2021}.

\section*{Acknowledgements}

We acknowledge fruitful discussion with Boris Gutkin, Nick Jones and
Michael Wilkinson. We are particularly grateful to Jacobus
J.M. Verbaarschot for helpful remarks on topological aspects of
Quantum Chromodynamics.  This work was funded by the German--Israeli
Foundation within the project \textit{Statistical Topology of Complex
  Quantum Systems}, grant number GIF I-1499-303.7/2019.

%
%
\bibliographystyle{unsrt}
\bibliography{lit}
%


\end{document}